# ExaGridPF: A Parallel Power Flow Solver for Transmission and Unbalanced Distribution Systems

Bin Wang, John Bachan, Cy Chan
Lawrence Berkeley National Laboratory
{wangbin, jdbachan, cychan}@lbl.gov

*Abstract*—This paper investigates parallelization strategies for solving power flow problems in both transmission and unbalanced, three-phase distribution systems by developing a scalable power flow solver, ExaGridPF, which is compatible with existing high-performance computing platforms. Newton-Raphson (NR) and Newton-Krylov (NK) algorithms have been implemented to verify the performance improvement over both standard IEEE test cases and synthesized grid topologies. For three-phase, unbalanced system, we adapt the current injection method (CIM) to model the power flow and utilize SuperLU to parallelize the computing load across multiple threads. The experimental results indicate that more than 5 times speedup ratio can be achieved for synthesized large-scale transmission topologies, and significant efficiency improvements are observed over existing methods for the distribution networks.

## I. Introduction

Solving power flow is the backbone for many applications in power system analysis, such as planning, operation, transient analysis, and contingency analysis. The objective of solving power flow is to determine the voltage profile, the current flow, and power loss, etc., across the power grid topology. Previous research has focused on modeling approaches [1], [2] in different systems and convergence properties of different algorithms, using multi-core processors or graphics processing units (GPU) [3]. As smart grid technologies advance, the power system is becoming more observable, enabling modeling and planning over larger networks with more details. The increased penetration of distributed energy resources (DERs) with their associated intermittency and uncertainty [4], such as electric vehicles (EVs) [5]–[9], solar PV generation [4], battery energy storage system (BESS) [4], [10], etc, will require operators to solve the power flow problem over large-scale grid topologies much more frequently to minimize grid constraint violations.

Essentially, the power flow problem solves the nonlinear power system equations due to the nonlinear properties of the ZIP load model [11], which is widely utilized by many power flow applications. Iterative algorithms have been proposed to solve the nonlinear equations, such as Gauss-Seidel [12], Newton-Raphson [12], Forward-Backward-Sweep [11], Z-Bus method [2], etc. According to [12], Newton's method is numerically superior to the Gauss-Seidel method and less prone to divergence for ill-conditioned problems, because of its quadratic convergence properties. For transmission systems, single line models are often utilized because most transmission lines are balanced, so they can be modeled using a simplified single-phase, balanced model. Voltage magnitude and angle are utilized to model the single-phase system. On the other hand, three-phase, unbalanced networks are very common in the distribution system, where the single-phase, balanced model assumption is no longer valid for the power flow calculation.

Open-source modeling tools have been developed to solve the power flow problems on different platforms, such as GridLab-D [13], OpenDSS [14], MAT-POWER [15], etc. Those applications are mostly desktop-based with limited computing resources. Newton-Raphson power flow algorithms have been implemented in Gridlab-D based on current injection method (CIM) using SuperLU [16] for solving sparse linear equations. In addition, three-phase Forward-Backward-Sweep (FBS) has been implemented in Gridlab-D with fast convergence performance. OpenDSS also works on desktop environments developed in the Delphi programming language, while MAT-POWER is developed in the MATLAB environment. However, none of these platforms can fully exploit the capability of the existing HPC platforms, such as Cori and Edison [17] at LBNL.

This paper investigates some parallelization strategies of solving power flow problems in both transmission and unbalanced distribution systems. This paper serves to 1) introduce the open-source implementation of ExaGridPF, which is a large-scale power flow solver designed for use on HPC platforms, 2) explore the parallelization strategies to speed/scale up the iterative power flow algorithms using multiple threads on multi-core computers, 3) present the experimental results and performance improvement over standard IEEE test cases and synthesized larger ones. This paper is organized as follows: Section II summarizes the formulations the power flow algorithms for transmission and distribution systems, respectively; Section III discusses the parallelization strategies of iterative algorithms using existing computing architectures; Section IV presents the experimental results and discussions; And Section V concludes the paper with future work.

This paper is sponsored by the LDRD funding by Lawrence Berkeley National Laboratory

## II. POWER FLOW ALGORITHMS

Solving power flow problem is required by many power system analysis and applications to determine voltage magnitude and angles (in each phase) across the power grid topology, as well as the current flow in each transmission line, etc., given the active ($P$) and reactive ($Q$) power drawn or injected into phase(s) at each bus. Due to the nonlinear properties of load modeling, i.e. ZIP load, iterative methods are needed to solve the nonlinear problem. Numerical approximation methods have been proposed to accelerate the computation and guarantee the convergence. Since most transmission systems are balanced, single-phase balanced models are suitable for the analysis, while the current injection method has been developed for unbalanced distribution systems with more details. The node-voltage representation is commonly used, which holds true for both the single-phase and three-phase algorithms.

$$I_{bus} = Y_{bus} \cdot V_{bus} \tag{1}$$

$$Y_{ii} = \sum_{j=0}^{n} y_{ij} \tag{2}$$

$$Y_{ij} = Y_{ji} = -y_{ij} \tag{3}$$

where $I_{bus}$ and $V_{bus}$ are the nodal current and voltage vectors, respectively. $Y_{bus}$ denotes the bus admittance matrix. The diagonal elements in $Y_{ii}$ are given by equation (2), while the off-diagonal elements in $Y_{bus}$ are given by (3), where $y_{ij}$ is the line admittance from bus $i$ to bus $j$.

### A. Single-phase power flow modeling

The polar form is adopted in the single-phase modeling of power flow algorithms, where the current for a specific bus $i$ can be expressed by equation (4), where $\theta_{ij}$ is the angle different between bus $i$ and bus $j$. $\delta_i$ and $\delta_j$ are the phase angles of bus $i$ and bus $j$, respectively. The complex form of power at bus $i$ is expressed in (5); Thus, the real and imaginary parts can be expressed separately by (6) and (7):

$$I_i = \sum_{j=1}^{n} |Y_{ij}||V_j| \angle \theta_{ij} + \delta_j \tag{4}$$

$$P_i - j * Q_i = V_i^* * I_i = |V_i| \angle -\delta_i \sum_{j=1}^{n} |Y_{ij}||V_j| \angle \theta_{ij} + \delta_j \tag{5}$$

$$P_i = \sum_{j=1}^{n} |V_i||V_j||Y_{ij}| \cos(\theta_{ij} - \delta_i + \delta_j) \tag{6}$$

$$Q_i = -\sum_{j=1}^{n} |V_i||V_j||Y_{ij}| \sin(\theta_{ij} - \delta_i + \delta_j) \tag{7}$$

Based on the above equations, the Jacobian matrix, which represents the linear relationship between small changes of $P_i$ and $Q_i$, i.e. $\Delta P_i$ and $\Delta Q_i$, and the voltage angle and magnitude deviations, i.e. $\Delta \delta_i$ and $\Delta |V_i|$, will be constructed as follows:

$$\Delta P_i = P_i^{sp} - P_i^{calc,} \tag{8}$$

$$\Delta Q_i = Q_i^{sp} - Q_i^{cal} \tag{9}$$

$$\begin{bmatrix} \Delta P \\ \Delta Q \end{bmatrix} = \begin{bmatrix} J_1 & J_2 \\ J_3 & J_4 \end{bmatrix} \begin{bmatrix} \Delta \delta \\ \Delta |V| \end{bmatrix} \tag{10}$$

Each element of the Jacobian can be obtained by taking the first-order derivatives of $P$ and $Q$ with respect to the corresponding voltage angle and magnitude deviations. In the runtime, both sides of the equation (10) and corresponding elements in the Jacobian for buses with power consumption or injection. The iterative process is illustrated as follows:

| | Single-phase Newton-Raphson Iteration |
|---|---|
| 1 | Iteration $k = 0$; |
| 2 | **Do** |
| 3 | $\quad k = k + 1$; |
| 4 | $\quad$ Calculate power mismatch by equation (8), (9) |
| 5 | $\quad$ Update Jacobian and compute $\Delta \delta_i^k$ and $\Delta |V_i^k|$ by (10); |
| 6 | $\quad \delta_i^{k+1} = \delta_i^k + \Delta \delta_i^k$ , $|V_i^{k+1}| = |V_i^k| + \Delta |V_i^k|$ ; |
| 7 | **While** $\max_{i \in N} |\Delta \delta_i^k| > \underline{err_\delta}$ or $\max_{i \in N} |\Delta |V_i^k|| > \underline{err_V}$ |

### B. Three-phase power flow (Current Injection Method)

Unlike the single-phase representation of the transmission lines, the three-phase system is more complicated with the addition of phase information. The mutual admittance of each branch is represented by a 3×3 matrix, taking into account the impact of phases over each other. The details of line models can be found in [1], [11]. For three-phase systems, we use the current injection method to formulate the power flow problem based on the admittance matrix construction methods (CIM) [18] discussed in [1], [11]. The admittance matrix will contain entries for each phase. In this paper, we use the indexing strategies in [1], where each available phase will be indexed by the bus number $i$ and the phase order $\phi$ in $\Phi_i$. For instance, the index of phase $a$ on bus $n$ will be $3 \cdot (i - 1) + 1$. Similarly, for phase $b$ of bus $i$, the index will be $3 \cdot (i - 1) + 2$. Thus, the individual element with bus $i$ and bus $j$ in the admittance matrix is represented as:

$$Y_{ij} = \begin{bmatrix} G_{ij}^{aa} + jB_{ij}^{aa} & G_{ij}^{ab} + jB_{ij}^{ab} & G_{ij}^{ac} + jB_{ij}^{ac} \\ G_{ij}^{ba} + jB_{ij}^{ba} & G_{ij}^{bb} + jB_{ij}^{bb} & G_{ij}^{bc} + jB_{ij}^{bc} \\ G_{ij}^{ca} + jB_{ij}^{ca} & G_{ij}^{cb} + jB_{ij}^{cb} & G_{ij}^{cc} + jB_{ij}^{cc} \end{bmatrix} \tag{11}$$

The specified active and reactive power for each bus, i.e. $P_i^{s,sp}$ and $Q_i^{s,sp}$, can be obtained by subtracting the load part $P_{gi}^s$ from the power output of generator $P_{li}^s$ in (12) and (13). Note that $P_{gi}^s$ is $0$ if there's no generator attached to such bus. The ZIP load model is given by (14) and (15).

$$P_i^{s,sp} = P_{gi}^s - P_{li}^s \tag{12}$$

$$Q_i^{s,sp} = Q_{gi}^s - Q_{li}^s \tag{13}$$

$$P_{li}^s = P_{P,i}^s + P_{I,i}|V_i| + P_{Z,i}^s|V_i|^2 \tag{14}$$

$$Q_{li}^s = Q_{P,i}^s + Q_{I,i}|V_i| + Q_{Z,i}^s|V_i|^2 \tag{15}$$

$$\Delta I_i^s = I_i^{s,sp} - I_i^{s,calc} \tag{16}$$

$$I_i^{s,sp} = \left(\frac{P_i^{s,sp} + j * Q_i^{s,sp}}{V_i^s}\right)^* \tag{17}$$

$$I_i^{s,calc} = \sum_{j \in \Omega_i} \sum_{t \in \alpha_k} Y_{ij}^{st} V_j^t \tag{18}$$

$$\Delta I_i^c = \left[\Delta I_{Img}^{a,i}, \Delta I_{Img}^{b,i}, \Delta I_{Img}^{c,i}, \Delta I_{Re}^{a,i}, \Delta I_{Re}^{b,i}, \Delta I_{Re}^{c,i}\right]^T \tag{19}$$

$$\Delta V_i^c = \left[\Delta V_{Re}^{a,i}, \Delta V_{Re}^{b,i}, \Delta V_{Re}^{c,i}, \Delta V_{Img}^{a,i}, \Delta V_{Img}^{b,i}, \Delta V_{Img}^{c,i}\right]^T \tag{20}$$

The current mismatches are calculated by equation (16), where $P_i^{s,sp}$ and $Q_i^{s,sp}$ are the specified active and reactive powers for a given phase $s$ at bus $i$. Accordingly, the specified current is given by (17). Due to the Kirchhoff's circuit laws (KCL), the current injected into bus $i$ should be equal to the summation of currents that flow out from this bus. Therefore, the alternative way of computing the injected current is to calculate the summation of the current flowing out into other buses (phases) from phase $s$, by equation (18). Due to the complexity of having the relation between the power (active and reactive) and voltage profile (magnitude and angles) in three-phase systems, current-voltage relation in equation (1) represented by the admittance matrix $Y$, is augmented to build the Jacobian instead. However, in the current injection method [18], real and imaginary parts of current and voltage on each bus are concatenated so that both LHS and RHS of the Jacobian are real vectors. Therefore, the concatenated changes of current and voltage vectors are given in equation (19) and (20).

Consequently, each block in Jacobian, which corresponds to one bus, is converted from a 3×3 matrix to 6×6. The updated block is represented by (21), where $B_{ij}$ and $G_{ij}$ are 3×3 matrice of real and imaginary parts of $Y_{ij}$. Therefore, the augmented Jacobian matrix with doubled size is given by (22).

$$J_{ij} = \begin{bmatrix} B_{ij} & G_{ij} \\ G_{ij} & -B_{ij} \end{bmatrix} \tag{21}$$

$$\begin{bmatrix} \Delta I_1 \\ \Delta I_2 \\ \vdots \\ \Delta I_N \end{bmatrix} = \begin{bmatrix} J_{11} & J_{12} & \cdots & J_{1N} \\ J_{21} & J_{22} & \cdots & J_{2N} \\ \vdots & \vdots & J_{kk}^* & \vdots \\ J_{N1} & J_{N2} & \cdots & J_{NN} \end{bmatrix} \cdot \begin{bmatrix} \Delta V_1 \\ \Delta V_2 \\ \vdots \\ \Delta V_N \end{bmatrix} \tag{22}$$

However, due to the specified active and reactive power consumption/injection on specific buses, the Jacobian is obtained by updating the diagonal blocks, equation (23) with adjusting elements for each bus, i.e. $a_i, b_i, c_i, d_i$, which are 3×3 matrice, depending on the ZIP load model. The adjusting elements represent the first-order linear relation between small changes of augmented current and voltage elements, given the active $P$ and reactive power $Q$ consumed at specific buses. The approaches to generate the adjusting elements are explicitly discussed in the appendix section of [18]. Since the primary purpose of this paper is to explore the computational aspects, we have not yet added power injection from distributed generations, i.e. PV buses.

$$J_{kk}^* = \begin{bmatrix} B_{kk} - a_k & G_{kk} - b_k \\ G_{kk} - c_k & -B_{kk} - d_k \end{bmatrix} \tag{23}$$

During the runtime, the current mismatch is evaluated, which is similar to the approach for the single-phase system, and then the voltage mismatch $\Delta V_i^c$ will be subsequently computed by solving the linear equation, $\Delta I^c = J \cdot \Delta V^c$. The iterative solving process stops when the maximum voltage deviation falls below a minimal threshold. The iterative process is as follows:

| | **Three-phase Newton-Raphson (CIM) Iterations** |
|---|---|
| 1 | Iteration $k = 0$; |
| 2 | **Do** |
| 3 | $\quad k = k + 1$; |
| 4 | $\quad$ Obtain $\Delta I_i^c$ by (12) – (20) |
| 5 | $\quad$ Update $J$ and compute $\Delta V_i^c$ by solving equation (22); |
| 6 | $\quad V_i^c = V_i^c + \Delta V_i^c$ ; |
| 7 | $\quad$ Convert $V_i^c$ into complex form $V_i$ ; |
| 8 | **While** $\max_{i \in N}|\Delta V_i^c| > \underline{err_V}$ |

## III. PARALLELIZATION STRATEGIES OF EXAGRIDPF

We investigate parallelization of the power flow solver using two techniques. Both solvers utilize the same outer Newton iteration, while using different strategies to solve the inner Jacobian system.

### A. Newton-Raphson with SuperLU Direct Linear Solver

We leverage the SuperLU [16] sparse direct solver library to investigate parallelizability of the power flow solver over multiple threads. There are two stages of the solver: factorization of the Jacobian into lower and upper triangular matrices, and solving the two resulting triangular systems. Care must be taken to permute the matrix to reduce the amount of fill-in that results during factorization, which can dramatically increase the number of non-zeros in the sparse matrix. We utilize the approximate minimum degree reordering from the COLAMD [19] package to limit the fill-in.

Since this is a direct method, there is no iteration necessary. However, the total amount of computation may still be high due to the inherent cost of the matrix factorization. The main benefit of using the slower direct method is that it is more robust for ill-conditioned matrices, allowing the power flow to be solved in cases where the Newton-Krylov method fails to converge.

### B. Newton-Krylov with Iterative Linear Solver

For the Newton-Krylov solver, we utilize an iterative method to solve the Jacobian system during each Newton iteration. Specifically, we tested a bi-conjugate gradient stabilized Krylov method [20], where each computational routine required is parallelized over multiple threads. Each computational thread is assigned a block of contiguous values in the vector being updated. For example, one of the main computations in a Krylov method is multiplication of the sparse Jacobian with a dense vector. In the sparse matrix vector multiplication, each thread reads a block of rows in the sparse matrix to update its assigned block of values in the vector, which can be calculated independently. Since no matrix factorization is required, this method can be very efficient so long as the matrix is well conditioned to ensure convergence of the iteration in a small number of steps.

One drawback of this method is that it does not guarantee convergence for ill-conditioned problems. For example, we observed non-convergent behavior in our replicated three-phase distribution problem. For these scenarios, the use of the direct SuperLU solver still allowed us to solve the power flow problem, though at increased computational cost. As with the Newton-Raphson method, we permute the Jacobian matrix, but the reason to do so is different since we are not factoring the matrix. By permuting the sparse matrix so that the non-zero elements are close to the diagonal, threads access and reuse elements of the vector that are close to one another, increasing the temporal and spatial data locality of access for each thread. This enhances the data access performance of the machine by more effectively utilizing the processor caches.

## IV. RESULTS AND DISCUSSION

We ran multiple experiments for both the transmission and distribution systems to evaluate the computational performance of ExaGridPF. We present some speed up metrics for running the core Newton-Krylov solver on single-phase systems, as well as a comparison of the execution time of the three-phase unbalanced distribution system power flow solver utilizing the Newton-Raphson solver with the Z-Bus method [1] implemented in the MATLAB environment.

For the single-phase transmission system model, we constructed large-scale cases by replicating the IEEE-30 test case with multiple generator buses across the topologies. To connect replicated blocks, we inserted random links to connect the replicated transmission network blocks. For instance, the buses of the second block are indexed from #31 to #60, and a link to connect bus number #60 to bus #29 may be created as there is an existing link between bus #29 and bus #30 in the original topology. Similar approaches to create large-scale topologies by replicating standard test cases with randomization can be found in previous research [21]. Therefore, circles (closed loops) can be found in the replicated topology, making the solving task challenging for the direct methods, such as Forward Backward Sweep or Z-Bus.

For each topology size, we evaluate the performance with different numbers of threads on a single compute node of the Cori computer at NERSC. The results are illustrated in Figure 1 (note the logarithmic Y-axis). The Krylov solver for the synthesized grid topology that has 48,000 grid buses takes 1.2 seconds with 1 thread and 210 milliseconds with 32 threads. Using the single thread performance as the baseline, we calculate the speedup ratio for each experiment by comparing the solver time with the baseline, and show the results in Figure 2. As indicated by figure, more than 5 times speedup can be achieved for our largest test case in the parallelized computational routine over multiple threads.

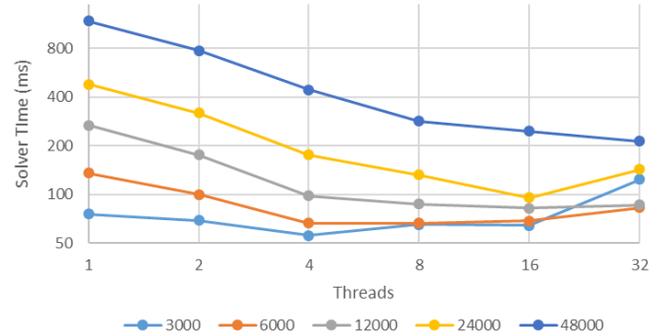

Figure 1 Krylov solver time versus thread count

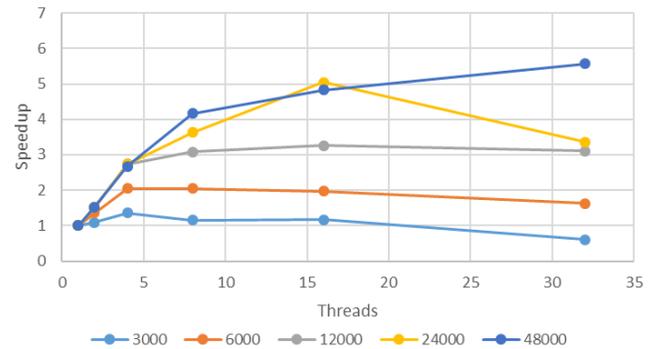

Figure 2 Krylov solver parallel speedup

Table 1 Solving time results from ExaGridPF and Z-Bus in Matlab

| Bus NO. | Z-Bus | | ExaGridPF | |
| --- | --- | --- | --- | --- |
| | Win i7/16GB | Mac i5/8GB | Cori (1 node) | |
| | | | Solving | Other |
| 0.9K | 1.1217 | 1.4113 | 0.319107 | 0.047195 |
| 1.8K | 3.9944 | 5.4393 | 0.389616 | 0.059578 |
| 3.6K | 17.6397 | 38.928 | 0.799942 | 0.092241 |
| 7.2K | × | × | 1.12324 | 0.218415 |
| 14.4K | × | × | 2.31822 | 0.450141 |
| 28.8K | × | × | 4.9977 | 0.954925 |
| 57.6K | × | × | 11.8052 | 2.047031 |
| 115K | × | × | 26.0988 | 4.346155 |
| 230K | × | × | 55.291 | 9.980889 |

For the three-phase, unbalanced distribution systems, the replication is based on IEEE 906 LV test case. We scale up the grid topology by replicating all buses after the first bus, which is connected to the substation transformer, whose parameters are adjusted to prevent a voltage drop constraint violation. We carried out a number experiments with varying number of replicated buses to evaluate the solver performance. For comparison, we evaluated the Z-Bus method in MATLAB on a Windows and Mac computer. The total solver times are shown in Figure 3. The ExaGridPF solver time increases roughly linearly with the number of grid buses, while the solving time of the benchmark solver (Z-Bus) exponentially increases, which could be caused by many factors. For the largest three-phase distribution system test case, we reduce the total solving time to less than 1 minute, as indicated by the results in Table 1. Time used for other miscellaneous operations, including Jacobian initiation, Jacobian update, intermediate parameters update, etc., are recorded in the "Other" column in Table 1.

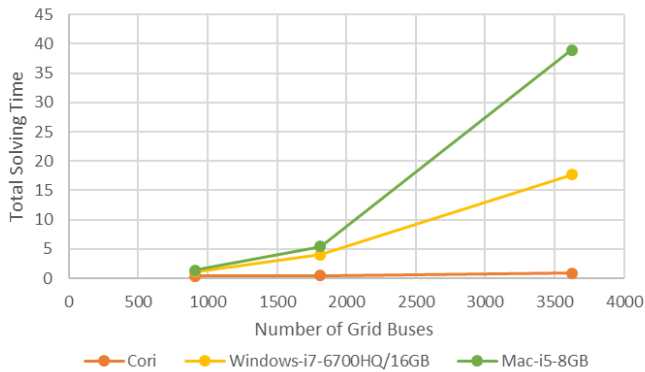

Figure 3 Newton-Raphson solver time versus Z-Bus method

However, as the problem size (block number) increases, the Z-Bus solver cannot proceed due to memory shortage. × in Table 1 indicates the lack of memory to perform the Z-Bus calculation. In contrast, ExaGridPF scales very well, which enables the user to perform more frequent power flow evaluations required by many trending applications in smart grid power systems analysis.

V. CONCLUSION

The future work will be to further explore the strategies to parallelize computations, especially for distribution systems with high fidelity. The results shown in this paper are for shared-memory computation. We are currently investigating the use of distributed-memory computation for solving power flow calculations over multiple nodes.

VI. ACKNOWLEDGEMENT

The research that leads to the results in this paper has been supported by LDRD funding at Computational Research Division (CRD), Lawrence Berkeley National Laboratory for the ExaGrid project.